\documentclass[conference]{IEEEtran}
\IEEEoverridecommandlockouts
\usepackage{cite}
\usepackage{amsmath,amssymb,amsfonts}
\usepackage{algorithmic}
\usepackage{graphicx}
\usepackage{textcomp}
\usepackage{xcolor}
\usepackage{fancyhdr}

\usepackage{booktabs}

\makeatletter
\def\ps@IEEEtitlepagestyle{
  \def\@oddfoot{\mycopyrightnotice}
  \def\@evenfoot{}
}
\def\mycopyrightnotice{
  {\hfill \footnotesize 978-1-6654-4238-1/21/\$31.00 \copyright 2021 IEEE \hfill}
}
\newcommand{\linebreakand}{ 
  \end{@IEEEauthorhalign}
  \hfill\mbox{}\par
  \mbox{}\hfill\begin{@IEEEauthorhalign}
}
\makeatother

\author{\IEEEauthorblockN{Chris Kottmyer}
\IEEEauthorblockA{\textit{student}\\
\textit{Harvard University} \\
ChrisKottmyer@gmail.com}
\and
\IEEEauthorblockN{Kevin Zhao}
\IEEEauthorblockA{\textit{student}\\
\textit{Harvard University} \\
zhaokevinusa@gmail.com}
\and
\IEEEauthorblockN{Zona Kostic}
\IEEEauthorblockA{\textit{John A. Paulson School of} \\
\textit{Engineering and Applied Sciences} \\
\textit{Harvard University}\\
zonakostic@seas.harvard.edu}
\linebreakand
\IEEEauthorblockN{Aleksandar Jevremovic}
\IEEEauthorblockA{\textit{Singidunum University} \\
ajevremovic@singidunum.ac.rs}
}

\def\BibTeX{{\rm B\kern-.05em{\sc i\kern-.025em b}\kern-.08em
    T\kern-.1667em\lower.7ex\hbox{E}\kern-.125emX}}

\begin{document}

\title{Roomsemble: Progressive web application for intuitive property search \\

}

\maketitle

\begin{abstract}
A successful real estate search process involves locating a property that meets a user’s search criteria subject to an allocated budget and time constraints.  Many studies have investigated modeling housing prices over time.  However, little is known about how a user’s tastes influence their real estate search and purchase decisions.  It is unknown what house a user would choose taking into account an individual’s personal tastes, behaviors, and constraints, and, therefore, creating an algorithm that finds the perfect match.  In this paper, we investigate the first step in understanding a user's tastes by building a system to capture personal preferences.  We concentrated our research on real estate photos, being inspired by house aesthetics, which often motivates prospective buyers into considering a property as a candidate for purchase.  We designed a system that takes a user-provided photo representing that person's personal taste and recommends properties similar to the photo available on the market.  The user can additionally filter the recommendations by budget and location when conducting a property search.  The paper describes the application’s overall layout including frontend design and backend processes for locating a desired property.  The proposed model, which serves as the application’s core, was tested with 25 users, and the study’s findings, as well as some key conclusions, are detailed in this paper.

\end{abstract}


\section{Introduction}
One of the last domains to be disrupted by technology is real estate, with research mainly focusing on price predictions \cite{zona_sol}, \cite{tmm}, \cite{location}, and \cite{modeling}. Although price is one of the most important factors in making a decision, it is usually not the only factor that motivates buyers. On the contrary, every transaction in the real estate market begins with a search for a desired appearance (except in the case of flipping, although even then the appearance of the house is very much taken into account). This often manifests in browsing real estate photos on websites, which is typically the first and often most important step when researching a property and eventually making a purchasing decision.  

In this paper, we present an intuitive method for searching real estate properties.  We focus on a user's tendency to browse images as a search criterion during the real estate search process. Buyers may be interested in more than one location, and services such as REX\textregistered, Zillow\textregistered, or Redfin\textregistered~allow geographic-based searches. Users rely on what interests them first - the appearance - and later adjust their search based on attributes such as location, square footage, and budget (for example, a user may prefer houses similar to a house appearing in a magazine). In our approach, the viewer can upload a picture similar to their personal preferences, and explore units with a similar appearance.  These similar units are recommendations from a machine learning model. In this manner, the model swifts through millions of images to provide the user with personalized recommendations, while the end user retains control of browsing the recommendations and making the final purchase decision. Every new search performed by the user could be used to retrain the machine learning algorithm and tailor the results to the user's specific tastes.
The method presented above is in the form of a responsive progressive web application that allows a user to either take a photo with their cell phone or upload an indoor or outdoor image of a residential property. In return, the user gets recommendations for similar units available on the market. The model was trained and tested on the Massachusetts real estate market. The application is targeted towards home buyers that are interested in finding houses that match their personal styles as represented by the uploaded photos. In addition, the model’s results were tested with 25 subjects. That way, we incorporate humans in the evaluation process and provide results focusing on correlations between human and model-based reasoning.

The rest of this work is organized as follows: we present related work in Section 2; image modeling techniques and the overall application architecture in Section 3; Section 4 describes the user study, metrics used and results obtained; we discuss the results and conclude the paper with suggestions for future work in Section 5.

\section{Related Work}

Many models have investigated the domain of image-based content retrieval due to the popularity of image content.   Content Based Image Retrieval (CBIR), for example, has been widely used  \cite{furuya2014visual} and it addresses the task of image recommendation based on a query (anchor) image. The problem with this approach, however, is that the sketch or image used includes single object of interest, whereas larger sets of images include many other elements (e.g., background) that are not presented with the anchor image. One of the solutions used by researchers was to apply the Visual Saliency Weighting to large sets of images in order to suppress clutter. Furthermore, the authors of this paper \cite{he2018content} investigate CBIR for images of general objects. The study compares the scale-invariant feature transform (SIFT) to two other feature descriptors and concludes that the SIFT outperforms other models. SIFT's experimental results were very promising, even for spatial images, and as a result, it was considered by our research team.

Convolutional Neural Networks (CNN) are well-known for their superior performance in image recognition and content retrieval. The paper \cite{yin2020image}, for example, can query similar images quickly due to the use of Local Sensitivity Hashing (which hashes high dimensional data and stores points in space close together in the hash table). The the two projects \cite{ullah2019image} and \cite{bookvis} suggest a fast image content retrieval and understanding for object-related data query, which are intuitively similar but applied in a different domain. The former project approaches modeling in two stages: first, it categorizes the query product, and then it recommends similar products. The latter project combines three different models, including SIFT and CNN, to retrieve data on objects based on a cell-phone taken image, regardless of the angle, crop, or quality of the image. All of the ideas mentioned above are used to some extent by our research team to develop the final model presented later in the paper.  CNN was specifically used to categorize pictures into specific room or outdoor image classes.  These categories allowed us to measure similarity within a category and use this to make recommendations for a photo belonging to a specific caetgory.

The research community is working hard to automate many processes in the real-estate domain. Such efforts, in particular, focus on applying image processing techniques \cite{cityforensics}, \cite{userpref}, \cite{law2018satellite}, \cite{imageappraisal}, and \cite{dlhouseprice}.
Another study \cite{de2014statistical}  employs a machine learning model approach to interpret floor plans. Although an intriguing concept, the main issue with floor plan analysis is that there is no standard notation, although, the authors discover that their pipeline is adaptable to almost all floor plans. In their original paper, the authors examine various descriptors, including SIFT, and discover that the blurred shape model performs best. While \cite{de2014statistical} primarily focuses on wall recognition, this paper  \cite{zeng2019deep} seeks to identify all components of a floor plan, including not only walls but also doors and room types (e.g. bedroom). The authors employ a deep multitask neural network, with one task predicting boundary components (e.g., wall, door/window) and another task predicting room type (e.g. bedroom, balcony). Furthermore, \cite{yamasaki2018apartment} uses a fully convolution network to segment floor plan images and then extract a graph model. The number of common edges between the graphs of two images can also be used to calculate the similarity between apartments.

There are other papers on floor plan retrieval, such as \cite{ozaki2019extraction}, which uses methods based on frequent subgraph mining to extract characteristics from floor plan images, or \cite{sharma2018rexplore}, which proposes a framework for retrieving similar floor plans based on the anchor image.
However, images (indoor, outdoor or satellite) are commonly used in the real-estate domain to extract characteristics and then combine them with other factors to forecast prices \cite{sharma2018rexplore} or properties demand \cite{tmm}. In this paper, we propose a method that combines multiple machine learning algorithms to deal with all of the edges and challenges (such as low-quality photos uploaded) and retrieve images of apartments with similar style, amenities, and/or structure. The chapters that follow go into greater detail about modeling as well as overall system architecture proposed by our research team.

\section{System Architecture}

We detail the cloud infrastructure components, user interface and progressive web application components used in our roomsemble application. Later sections cover use of Places365 CNN, Triplet Loss and SIFT models to categorize an image and provide recommendations for pictures uploaded by roomsemble end-user.

\begin{figure}[h!]
\vspace{-3mm}
    \includegraphics[width=\linewidth]{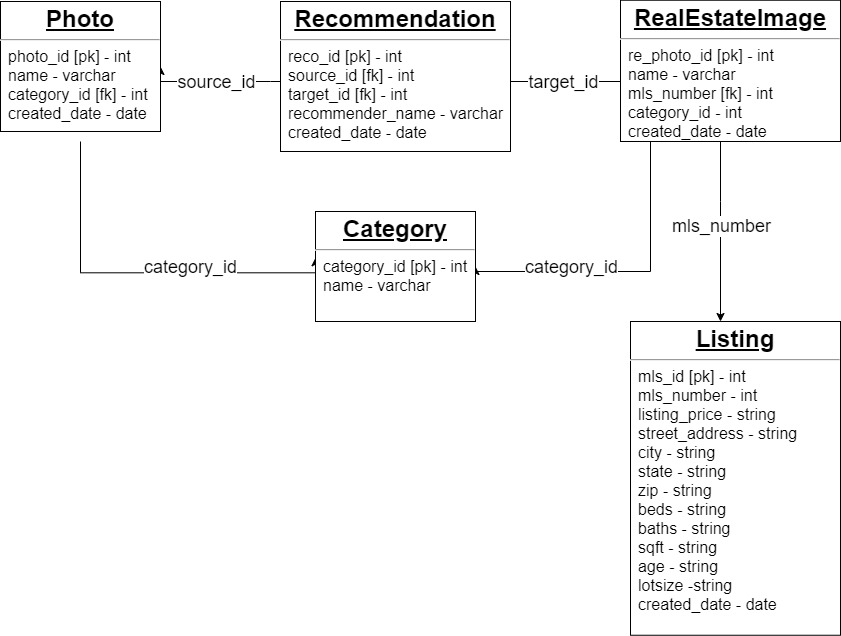}
    \caption{Database scheme.}
    \label{db}
\end{figure}

\subsection{Hosting and Infrastructure}

Roomsemble was prototyped on AWS services with a focus on Route 53, EC2 and EBS services.  We utilized a C4.xlarge EC2 instance running Ubuntu 18.04 operating system and attached two SSD-based EC2 volumes to it: root drive with 20 gb partition hosting the linux and an 80 gb secondary drive used by postgresql and as a filesystem to store images.  Roomsemble is based on a Flask application with gunicorn and Nginx set-up to serve Flask pages via WSGI.  We associated an elastic ip address to the EC2 instance and served the Roomsemble app over https port 443 with a self-signed SSL certificate certificate. We provide a SystemD unit file to start and stop the Roomsemble service.

\subsection{Database Architecture}

The application utilizes postgresql to support the Roomsemble application. The tables are: Photo - a user’s photos, Listing - MLS listing information, RealEstateImage - photos associated with MLS Listings, Category - list of photo categories and Recommendation - a table of recommendations associating Photo with RealEstateImage. Both Photo and RealEstateImage on EBS Volume with file names stored in both the Photo and RealEstateImage tables.  We pre-populate Listing, Photo and Category tables.  Photo and Recommendation table entries are generated when a user uploads a photo.

\begin{figure}[h!]
    \centering
    \includegraphics[width=150pt]{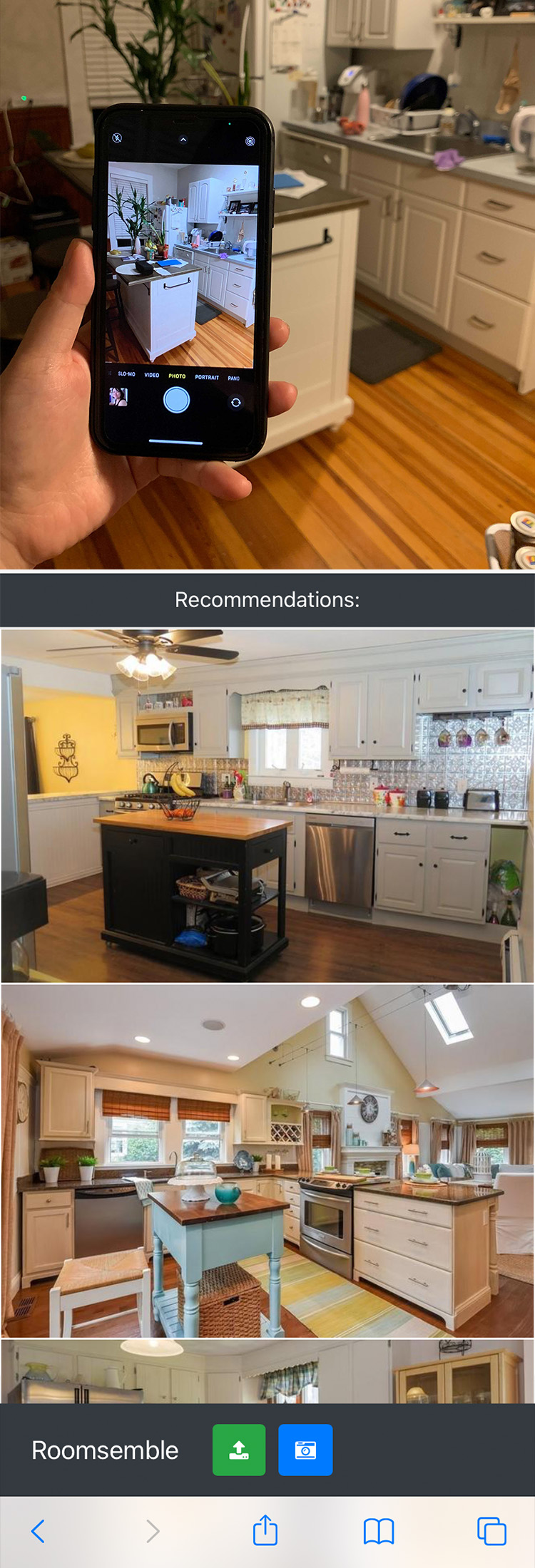}
    \caption{User Interface and Recommendations.}
    \label{ui}
\end{figure}


\subsection{User interface}

This user interface (Fig. \ref{ui}) presents the latest recommendations if a user has uploaded a photo.  If no photo has been uploaded, no recommendations appear.  Photo’s are uploaded by the green upload icon on the bottom right.  Alternatively, the blue take photo button can be used to engage a user’s camera to take a picture to be uploaded.

Once recommendations have been generated, a user can click on a photo recommendation to go to the MLS listing page.  This includes information on the age of the MLS Listing, street address, sales price, number of bedrooms, bathrooms, square foot and lot size.  The webpage also includes a photo gallery of pictures associated with the MLS listing, which when clicked on provide a navigable interface provided by lightbox.js. The order of the photo gallery is ranked using PCA to reduce the image features to one dimension.

The UI is designed with mobile first approach with 3 media query breakpoints: cell phone, tablet and Desktop respectively.  The layout is designed using a front-end framework: bootstrap and font-awesome.  The images are organized within a CSS Flexbox grid.  This allows us to seamlessly switch between 3 types of devices.

\subsection{PWA features}

Roomsemble is a progressive web application (PWA).  Progressive web applications can be installed on a cell phone's home page and have the ability to work without an internet connection.  We provide these features through a red download button on the homepage, which installs Roomsemble as an application on a user’s home page or desktop.  The offline capabilities are provided by service-worker.js file, which caches the most recent recommendations in the browser allowing it to be viewed in an offline capacity.

\subsection{Uploading Images and Machine Learning algorithms}

Recommendations are created by a 4-step process.  A user uploads an image either using the green upload or blue camera button.  The image is uploaded to the roomsemble server, resized and saved to disk.  The image is then fed to a categorization algorithm, which categorizes it.  The image and category are then sent to a recommendation algorithm, which filters for images within the category and then finds the 12 most similar images.  This is returned to the user as recommendations on the home page.  Below are more details about the categorization and recommendation algorithm.

\subsection{Categorization algorithm}

For categorizing images, we used Places365 pre-trained widerresnet algorithm based on the \cite{tmm}.  We categorized 366,361 real-estate photos into 10 categories.  This same pre-trained network is used to classify images the user uploads into the same 10 categories.  The category is used by the recommendation algorithm to filter and then rank the possible recommendations for the user’s photo.  More detail on this in the next section.


\section{Image Recommendation Model}

Our image recommendation algorithm is an ensemble of two different models.

\paragraph{Triplet Loss} 
Triplet loss is a common method used to learn image representations. 
As its name implies, the loss function operates on triplets of images, with each triplet consisting of an anchor image, positive image, and negative image. 
The triplets are created such that the anchor image is similar to the positive image but dissimilar to negative image. 
This loss function minimizes the Euclidean distance between the anchor image with the positive image while maximizing the distance between the anchor image and the negative image. 
The loss be rewritten as:
\begin{equation}
    \centering
    L(x_a, x_p, x_n) = \lVert f(x_a) - f(x_p) \rVert_2^2 - \lVert f(x_a) - f(x_n) \rVert_2^2 
\end{equation}
where $L$ denotes the triplet loss function, $f(x)$ is the embedding generated for image $x$, and $x_a$, $x_p$, and $x_n$ are the anchor image, positive image, and negative image respectively. 

\paragraph{Triplet Loss Sampling}
Triplet loss \cite{schroff2015facenet} is usually employed with labelled data. 
Positive images are sampled from the same class as the anchor image, while negative images are selected from a different class.
In our case with real estate, we can classify images using the places365 CNN. 
However, the places365 model can only classify images into broad categories like kitchens or bathrooms.
If we sample positive and negative images from these large categories, the model would only be able to distinguish between different types of rooms and unable to recognize the stylistic differences between rooms of the same category.
The style of a room is inherently subjective, and \cite{ataer2019room} categorize 800K images of house interiors into seven styles by asking 10 experts to tag each image. 
However, to the best of our knowledge, there are no publicly available datasets for room style. 
Instead, we use an automated process to generate our triplets, relying on an obvious principle: a room is similar to itself.
Our procedure is simple:
\begin{enumerate}
    \item Categorize all models using the places365 CNN 
    \item Find images with the same Multiple Listing Service (MLS) number that are classified as the the same category\footnote{We do not include bedrooms in our triplets as most houses have more than one bedroom.}
    \item Create triplets using the images with the same house and category as the anchor and positive images, while randomly sampling the negative image from images of the same category.
\end{enumerate}
For example, in step 2 we may find that House A has two images categorized as kitchens. 
Since most houses have only one kitchen, it is likely that the images are of the same kitchen but taken from a different angle.
Then, we can generate a triplet by using one of the images as the anchor image and the other image as the positive image, as well as a kitchen image from House B for the negative image. 

\begin{figure}[h!]
\vspace{-3mm}
    \centering
    \includegraphics[width=190pt]{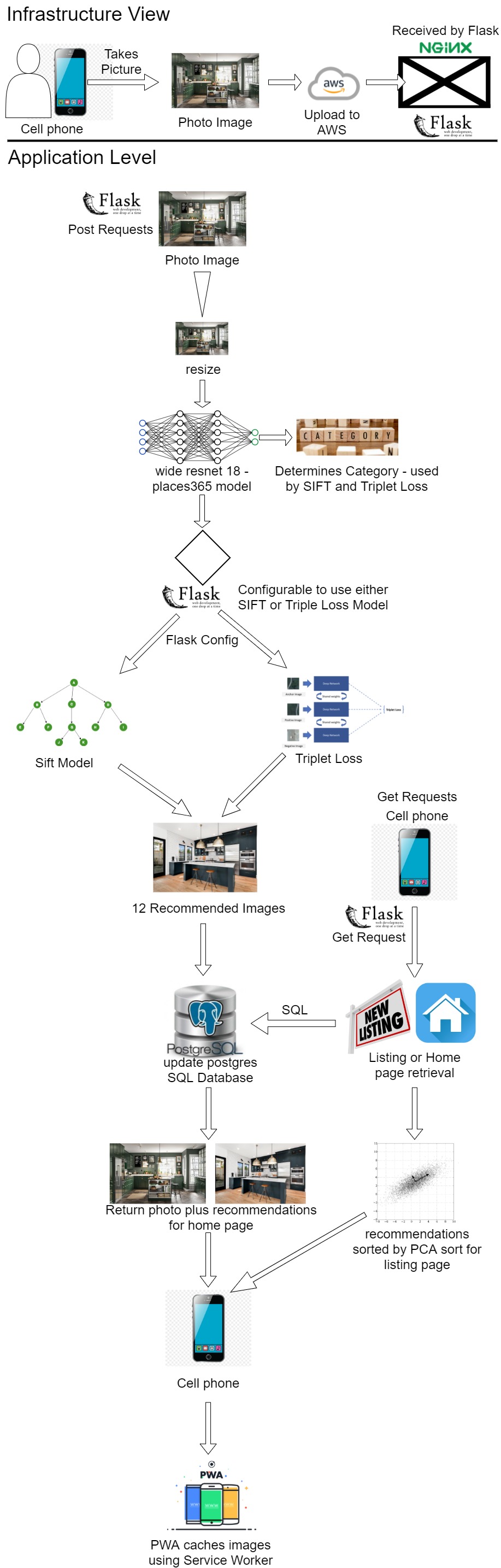}
    \caption{System Architecture.}
    \label{sys_architecture}
\end{figure}


\paragraph{Scale-invariant feature transform (SIFT)}

SIFT \cite{lowe1999object} is an algorithm used to extract features from images.
The resulting features are mostly unaffected by factors such as scaling, rotation, and illumination.
The SIFT algorithm identifies a set of locally distinct features, called keypoints, with a descriptor associated with each feature.
Ideally, images of the same object(s) from another angle or scale would share the same keypoints. 
Similar images would share a higher proportion of keypoints.
To determine whether a keypoint of an image is common to another image, we compute the Euclidean distances between the descriptor of the keypoint and all the descriptors of the other image.
If the ratio between the shortest distance and second shortest distance is greater than a threshold, then the keypoint is considered to be common between the two images.
The similarity score between two images is the ratio of the number of shared keypoints to the minimum number of keypoints. 
For example, if image A has 10 features and image B has 15 features, with 7 features determined to be shared between the images, the similarity score would be $\frac{7}{10}$.

\section{Evaluation}

\begin{figure*}
    \includegraphics[width=\textwidth]{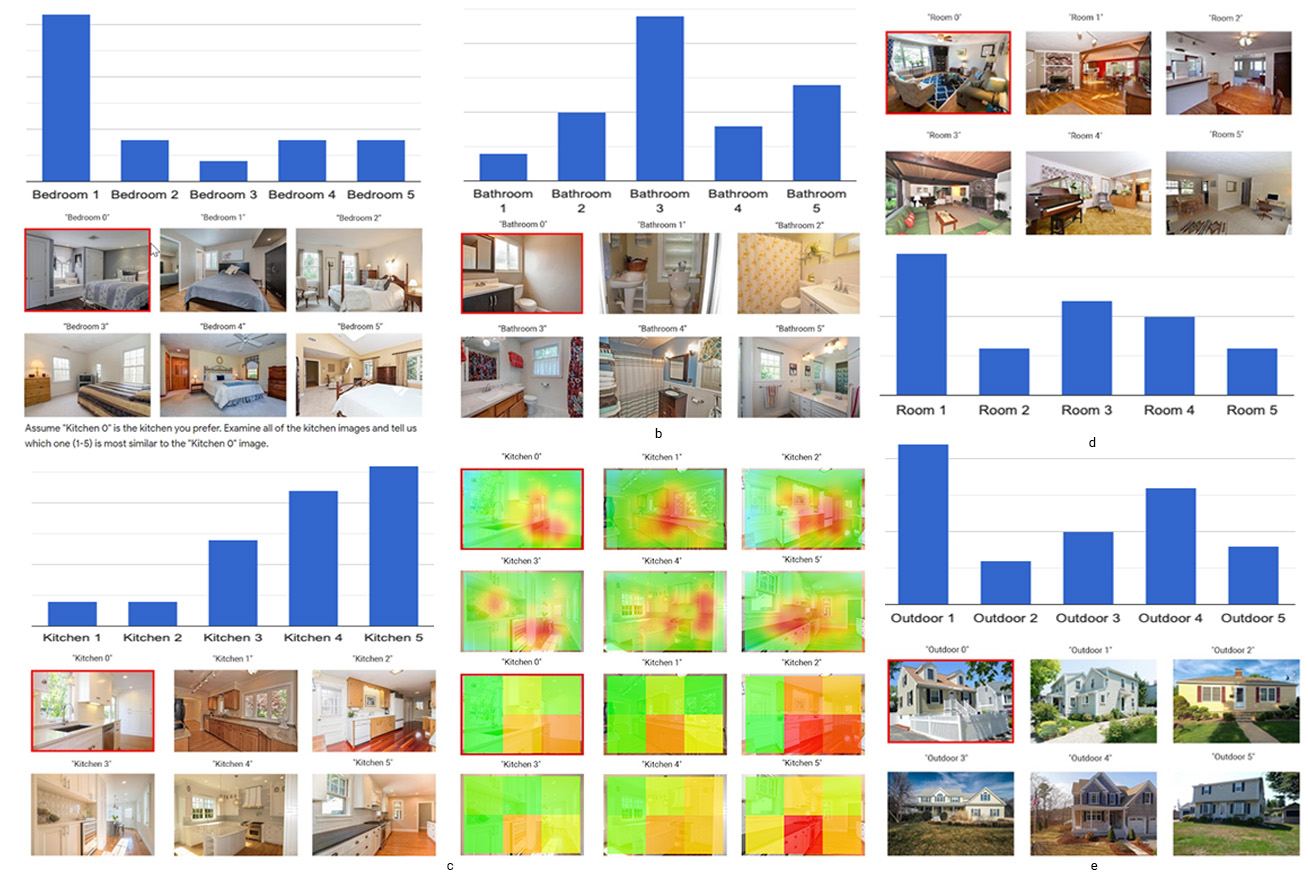}
    \caption{Survey conducted with 25 participants, asking participants to choose the most similar image to the anchor image in 5 different categories.}
    \label{survey_imgs}
\end{figure*}

We conducted a user study with 25 participants to better understand the usability of the application (specifically the machine learning models). The study focuses on non-domain experts in particular (general audience). We used five questions in which participants were asked to choose the image most similar to an "anchor" image (the query image or the image of interest) within a specific category such as kitchen or bathroom. We then compared the results of the users to the model's recommendations. We show the results of five categories (bedroom, bathroom, kitchen, living room, and outdoor) as well as the correlation between users’ and model’s preferences in this section of the paper.

Question 1 asks respondents to compare the anchor image (“Bedroom 0”) to the other 5 images and select one (or more) that appears to be the most similar to the anchor image. According to the results (Fig. \ref{survey_imgs}, a), the image with the highest user rating is “Bedroom 1,” which is also the image with the highest similarity score according to the algorithm.

Question 2 asked participants the same question, but focused on the "Bathroom" category. Although “Bathroom 3” received the most votes (from users and was the highest rated image by the model), it is worth noting that other bathroom images (2 and 5) were also highly rated by users and ranked second and third in similarity by the model. We made some interesting observations in the process of comparing results, such as the visual perception similarities between the users’ and model's choices. Both users and the model are focused on details (wide sink counter in images 0, 2, 3, and 5); image composition (perspective with images 0, 3, and 5); image brightness (0, 3, and 5); and the presence and position of windows (natural light). These factors played an important role in ranking images by style.

Question 3 asked participants to rank “Kitchen” images based on similarity. We discovered a disagreement in choices between users and the model leading to interesting results, which we further explored and will discuss below (it is important to be able to understand not only when the model makes a mistake, but also why). To begin, we see that the users and model differed in what they considered to be the most similar image (the model suggested "Kitchen 3," while users picked "Kitchen 5"). Surprisingly, “Kitchen 3,” the model's highest ranked kitchen image, is currently ranked third among users. The task was difficult for both the model and the users (images 3, 4, and 5 are very similar in style, composition, and other previously mentioned elements) and so we decided to run an additional analysis to understand those differences.

When comparing 2D information distribution, kitchens 1 and 3 share the most similarity with kitchen 0 - most information is located in the bottom-center and bottom-right regions. However, the color palette of kitchen 1, as well as the image perspective, are more similar to kitchen 0.

To begin, we create CAM activation maps similar to \cite{tmm} to see if there is a difference in attention. After discovering that attention maps produce similar results, we decided to apply the entropy technique developed by \cite{tmm} and discovered some exciting insights (Fig. \ref{survey_imgs}, c). Visual entropy is an excellent technique for understanding the spatial distribution that is most similar between "Kitchen 0" and "Kitchen 3." When we carefully examine the images, we see that the similarity in style is indeed greatest between these two images. Since our application focuses on replicating a user's preferences, the "Kitchen" example provides excellent feedback into the user's mental model and provides adjustments we can incorporate into our model to better approximate the user's preferences.


Question 4 asked participants to select the most similar images from a group of "Living room" or "Room" images. The most selected option (Room 1) was different from the model's top choice (Room 4). However, overall rankings were pretty accurate, as the model and humans both predicted Rooms 1, 3, 4 for the top 3 choices. Finally, participants were asked to compare "Outdoor" images. The users are in agreement with the model's choices, with "Outdoor 1" selected as the best by both.


\section{Conclusions and Future Work}

In this paper, we present the Roomsemble app, which recommends actively listed real estate properties based on a user’s stylistic preferences and personal taste. Since the PWA component and the ML model are the most novel aspects of this application, we concentrated our research on machine learning modeling, interpretability, and evaluation. While results were promising, this research identified problems and led to corresponding solutions that will be used in future works.  We want to gain a better understanding of the model and intend to scale it to a large number of users, while also increasing the number of images and categories available in the application.  During the scaling to a larger number of users, we will expand our survey analysis beyond the 25 individuals involved to increase the statistical significance of our findings as well as investigate further questions.

The app is currently set up to work on a single computer system, which limits its scope to a small group of users. Ideally, we would prefer if a user could walk into a random apartment, take pictures, and rate the suggestions on the spot. To enable such an application, we would have to utilize commercial approaches such as setting up multiple servers to scale out both the web server to handle more web requests and also database by using read-replicas and an external image server to scale out both the data and image retrieval services. That way we would be able to independently scale out each service based on its overall latency and CPU load. 

Lastly, we believe that the categorization algorithm and two recommendation algorithms could be further explored by looking into other models and potentially ensembling them with the current models. This could produce an even better recommendation system or provide more fine-tuned categories tailored specifically to the real-estate domain.

\bibliography{references}{}
\bibliographystyle{plain}

\end{document}